\documentclass{article}
\usepackage{amsmath,amsthm}
\usepackage{amssymb,latexsym}
\usepackage[mathscr]{eucal}
\usepackage{setspace}
\usepackage{graphics}
\usepackage{array}

\newcommand{\pa}{\partial}
\newcommand{\pr}{\prime}

\newcommand{\na}{\nabla}
\newcommand{\ti}{\times}

\newcommand{\lp}{\left(}
\newcommand{\rp}{\right)}
\newcommand{\lb}{\left[}
\newcommand{\rb}{\right]}
\newcommand{\lc}{\left\{}
\newcommand{\rc}{\right\}}

\newcommand{\lw}{\left|}
\newcommand{\rw}{\right|}
\newcommand{\be}{\begin{equation}}
\newcommand{\ee}{\end{equation}}

\newcommand{\ihat}{\bf\hat{i}}

\begin{document}

\title{\textbf{Beltrami States in Two-dimensional Electron Magnetohydrodynamics}}         
\author{B. K. Shivamoggi\\
University of Central Florida\\
Orlando, FL 32816-1364, U.S.A.\\
}        
\date{}          
\maketitle

\large{\bf Abstract}

In this paper, the Hamiltonian formulations along with the Poisson brackets for two-dimensional (2D) electron magnetohydrodynamics (EMHD) flows are developed. These formulations are used to deduce the Beltrami states for 2D EMHD flows. In the massless electron limit, the EMHD Beltrami states reduce to the force-free states, though there is no force-free Beltrami state in the general EMHD case.

\pagebreak

\noindent\Large\textbf{1. Introduction}\\

\large In the magnetohydrodynamics (MHD) model, the dynamics is dominated by ions with electrons serving to shield out rapidly any charge imbalances. In electron MHD (EMHD), with $\ell \ll \rho_{s_i}, ~s = i, e, \rho_s$ being the gyro-radius, on the other hand, the dynamics is dominated by electrons with the demagnetized ions serving to provide the neutralizing static background (Kingsep et al. \cite{Kin}, Gordeev et al. \cite{Gor}). The assumptions underlying the EMHD model are $\ell \ll d_i$, where $d_s \equiv c/\omega_{p_s}$ is the skin depth, and that the frequencies involved are greater than $\omega_{c_i}$ and $\omega_{p_i}$, $\omega_c$ being the cyclotron frequency. Some theoretical developments including two invariants for the 2D EMHD equations were given by Biskamp et al. \cite{Bis} .

On the other hand, equations governing several plasma dynamics models are known to admit a significant class of exact solutions (Shivamoggi \cite{Shi}) under the {\it Beltrami} condition - the local current density is proportional to the magnetic field - the {\it force-free} state (Lundquist \cite{Lun}, Lust and Schluter \cite{Lus}). These Beltrami solutions are known to correlate well with real plasma behavior (Priest and Forbes \cite{Pri}, Schindler \cite{Sch}). The purpose of this paper is to develop the Hamiltonian formulation along with the Poisson bracket for two-dimensional (2D) EMHD flows. This formulation is used to deduce the Beltrami states for 2D EMHD flows. These considerations are extended to the massless-electron limit (Huba and Rudakov \cite{huru}, Rheinhardt and Geppert \cite{rhgep}, Wood et al. \cite{wo}).

\vspace{.3in}

\noindent\Large\textbf{2. Governing Equations of 2D EMHD}\\

\large On assuming that the displacement current $\partial{\bf{E}}/\partial t$ is negligible, which is valid if \\$\omega\ll \omega_{p_e}^2/\omega_{ce},$ the continuty of the electron flow implies

\be\notag
\frac{\partial n_e}{\partial t} =-\nabla\cdot\lp n_e\bf{v}_e\rp=\nabla\cdot\lp\frac{\bf{J}}{e}\rp=\frac{c}{e}~\nabla\cdot\lp\nabla\times\bf{B}\rp=0
\ee

\noindent or

\be\tag{1}
n_e=const.
\ee

\large The 2D EMHD system of equations can then be written in terms of two scalar potentials (Biskamp et al. \cite{Bis}) - the magnetic flux function $A$ describing the in-plane magnetic field ${\bf B} = \na \ti A \ {\ihat}_z$ and the stream function $\psi$ describing the in-plane solenoidal electron flow velocity ${\bf v}_e = \na \ti \psi \ {\ihat}_z$, which is proportional to the in-plane current density (so $\psi$ also represents the out-of-plane magnetic field):\\
* the equation of generalized vorticity:

\be\tag{2}
\frac{\pa}{\pa t} \lp \omega + \frac{1}{d_e^2} \psi \rp + \lp {\bf v}_e \cdot \na \rp \omega - \frac{1}{m_e n_e c} \lp {\bf B} \cdot \na \rp J = 0
\ee

* the generalized Ohm's law:

\be\tag{3}
\frac{\pa}{\pa t} \lp A + \frac{d_e^2}{c} J \rp + \lp {\bf v}_e \cdot \na \rp \lp A + \frac{d_e^2}{c} J \rp = 0
\ee

\noindent where,

\be\tag{4}
\frac{1}{c} J = -\na^2 A ~, ~\omega = -\na^2 \psi.
\ee

Equations (1) and (2) may be rewritten as

\be\tag{5}
\frac{\pa \psi_e}{\pa t} = \lb \psi, \psi_e \rb + \frac{1}{c} \lb J, A_e \rb
\ee

\be\tag{6}
\frac{\pa A_e}{\pa t} = \lb \psi, A_e \rb
\ee

\noindent where,

\be\tag{7}
\left.
\begin{aligned}
\psi_e &\equiv m_e n_e \lp \psi/d_e^2 + \omega \rp\\
A_e &\equiv A + \lp d_e^2/c^2\rp J\\
\lb A, B \rb &\equiv \na A \ti \na B \cdot {\ihat}_z.
\end{aligned}
\rc
\ee

In the linear limit, on assuming a constant magnetic field ${\bf B}_0$ in the $x$-direction and a space-time dependence $e^{i \lp k x - \omega t \rp}$ for the perturbed quantities, equations (5) and (6) give the whistler waves ($\psi_k = \pm k A_k$ - the whistler condition),

\be\tag{8}
\omega \approx k^2 d_e V_{A_e}
\ee

\noindent which are dispersive unlike the Alfv\'{e}n waves. Here, $V_{A_e} \equiv B_0/\sqrt{m_e n_e}$.

\vspace{.3in}

\noindent\Large\textbf{3. The Hamiltonian Formulation}\\

\large Equations (5) and (6) have the Hamiltonian formulation (Biskamp et al. \cite{Bis}),

\be\tag{9}
H = \frac{1}{2} \int\limits_{\mathscr{A}} \lp \psi \psi_e + \frac{1}{c} A_e J \rp d \mathscr{A}
\ee

\noindent $\mathscr{A}$ being the area occupied by the plasma.

The Hamilton's equations are, on taking $(\psi_e, A_e)$ to be the canonical variables, given by

\be\tag{10}
\lp
\begin{matrix}
\displaystyle\frac{\pa \psi_e}{\pa t}\\
\\
\displaystyle\frac{\pa A_e}{\pa t}
\end{matrix}
\rp
= \mathscr{J} \lp
\begin{matrix}
\displaystyle\frac{\delta H}{\delta \psi_e}\\
\\
\displaystyle\frac{\delta H}{\delta A_e}
\end{matrix}
\rp
\ee

\noindent where $\mathscr{J}$ is a $\lp \psi_e, A_e \rp$-dependent differential operator which produces a skew-symmetric transformation of vector functions vanishing on $\pa \mathscr{A}$ and satisfies a closure condition on an associated symplectic two-form,

\be\tag{11}
\mathscr{J} \equiv \lp
\begin{matrix}
\lb \cdot, \psi_e \rb & \lb \cdot, A_e \rb\\
\\
\lb \cdot, A_e \rb & 0
\end{matrix}
\rp
\ee

\noindent and $\delta H/\delta q$ is the variational derivative. Equation (10) is just equations (5) and (6).

The operator $\mathscr{J}$ induces a Poisson bracket as follows,

\be\tag{12}
\begin{aligned}
\lb F, G \rb &\equiv \lp \lc \frac{\delta F}{\delta \psi_e}, \frac{\delta F}{\delta A_e} \rc, ~\mathscr{J} \lc \frac{\delta G}{\delta \psi_e}, \frac{\delta G}{\delta A_e} \rc^T \rp\\
&= \int\limits_{\mathscr{A}} \lc \frac{\delta F}{\delta \psi_e}, \frac{\delta F}{\delta A_e} \rc \lp \begin{matrix} \lb \displaystyle\frac{\delta G}{\delta \psi_e}, \psi_e \rb + \lb \displaystyle\frac{\delta G}{\delta A_e}, A_e \rb\\ \lb \displaystyle\frac{\delta G}{\delta \psi_e}, A_e \rb \end{matrix} \rp d \mathscr{A}\\
&= \int\limits_{\mathscr{A}} \lc \psi_e \lb \frac{\delta F}{\delta \psi_e}, \frac{\delta G}{\delta \psi_e} \rb + A_e \lp \lb \frac{\delta F}{\delta \psi_e}, \frac{\delta G}{\delta A_e} \rb - \lb \frac{\delta G}{\delta \psi_e}, \frac{\delta F}{\delta A_e} \rb \rp \rc d \mathscr{A}
\end{aligned}
\ee

\noindent which is a bilinear function defined on admissible functionals $F \lb \psi_e, A_e \rb$ and $G \lb \psi_e, A_e \rb$ satisfying the gauge condition,

\be\tag{13a}
\na \cdot \lp
\begin{matrix}
\displaystyle\frac{\delta F}{\delta \psi_e}\\
\\
\displaystyle\frac{\delta F}{\delta A_e}
\end{matrix}
\rp
= 0, ~\na \cdot \lp
\begin{matrix}
\displaystyle\frac{\delta G}{\delta \psi_e}\\
\\
\displaystyle\frac{\delta G}{\delta A_e}
\end{matrix}
\rp = 0 ~\text{in} ~\mathscr{A}
\ee

\noindent and the boundary conditions,

\be\tag{13b}
\lc \lw \frac{\delta F}{\delta \psi_e} \rw, \lw \frac{\delta F}{\delta A_e} \rw \rc, \lc \lw \frac{\delta G}{\delta \psi_e} \rw, \lw \frac{\delta G}{\delta A_e} \rw \rc = {\bf 0} ~\text{on} ~\pa \mathscr{A}.
\ee

The Poisson bracket (12) satisfies the anti-symmetry property,

\be\tag{14a}
\lb F, G \rb = -\lb G, F \rb
\ee

\noindent and the Jacobi identity,

\be\tag{14b}
\lb \lb F, G \rb, K \rb + \lb \lb G, K \rb, F \rb + \lb \lb K, F \rb, G \rb = 0.
\ee

In the absence of elecron inertia ($d_e \to 0$), (12) reduces to the MHD result given by Holm et al. \cite{Holm}.

\vspace{.3in}

\noindent\Large\textbf{4. Beltrami States}\\

\large The Casimir invariants for this problem are annihilators (with respect to any pairing functionals) of the Poisson brackets, which become degenerate when expressed in terms of these {\it natural} quantities. The Casimir invariants are therefore solutions of the equations,

\be\tag{15}
\mathscr{J} \lp
\begin{matrix}
\displaystyle\frac{\delta \mathscr{C}}{\delta \psi_e}\\
\\
\displaystyle\frac{\delta \mathscr{C}}{\delta A_e}
\end{matrix}
\rp = \lp
\begin{matrix}
0\\
\\
0
\end{matrix}
\rp.
\ee

It may be verified that two such solutions are,

\be\tag{16}
\mathscr{C}_{(1)} = \int\limits_{\mathscr{A}} f \lp A_e \rp d \mathscr{A}
\ee

\be\tag{17}
\mathscr{C}_{(2)} = \int\limits_{\mathscr{A}} \psi_e g \lp A_e \rp d \mathscr{A}
\ee

\noindent for any functions $f$ and $g$. (16) and (17) were deduced by Biskamp et al. \cite{Bis} by working directly with equations (5) and (6). $\mathscr{C}_{(1)}$ signifies the conservation of generalized magnetic flux in EMHD.

Minimization of $H$, keeping $\mathscr{C}_{(1)}$ fixed, gives

\be\tag{18}
\frac{\delta H}{\delta A_e} = \lambda_{(1)} \frac{\delta \mathscr{C}_{(1)}}{\delta A_e}
\ee

\noindent or

\be\tag{19}
\frac{1}{c} J = \lambda_{(1)} f^\pr \lp A_e \rp
\ee

\noindent which prescribes the Grad-Shafranov equilibria for 2D EMHD. There is no force-free state in 2D EMHD. This situation changes, however, in the massless-electron limit (see Section 5).

On the other hand, minimization of $H$, keeping $\mathscr{C}_{(2)}$ fixed, gives

\be\tag{20}
\frac{\delta H}{\delta \psi_e} = \lambda_{(2)} \frac{\delta \mathscr{C}_{(2)}}{\delta \psi_e}
\ee

\noindent or

\be\tag{21}
\psi = \lambda_{(2)} g \lp A_e \rp
\ee

\noindent or

\be\tag{22}
{\bf v}_e = \lambda_{(2)} g^\pr \lp A_e \rp {\bf B}_e \equiv a {\bf B_e}
\ee

\noindent where,

\be\tag{23}
{\bf B}_e \equiv \nabla \times A_e \ \hat{{\bf i}}_z = {\bf B} - d_e^2 \na^2 {\bf B}.
\ee

\noindent (21) describes the generalized Alfv\'{e}nic state in 2D EMHD.

\vspace{.3in}

\noindent\Large\textbf{5. EMHD in the Massless Electron Limit}\\

\large In the massless electron limit, the Ohm's law is

\be\tag{24}
{\bf{E}} + \frac{1}{c}~{\bf{v}}_e\times{\bf{B}}={\bf{0}}
\ee

\noindent which corresponds to the constraint of zero electric field in the reference frame moving with the electron fluid. \\

Using,

\be\tag{25}
{\bf{v}}_e=-\frac{\bf{J}}{n_e e}=-\frac{c}{n_e e}\nabla\times{\bf{B}}
\ee

\noindent and Faraday's Law,

\be\tag{26}
-\frac{1}{c}~\frac{\partial{\bf{B}}}{\partial t}=\nabla\times{\bf{E}}
\ee

\noindent equation (24) leads to (Huba and Rudakov \cite{huru}),

\be\tag{27}
\frac{\partial{\bf{B}}}{\partial t}+\frac{c}{n_e e}\nabla\times\lb\lp\nabla\times{\bf{B}}\rp\times{\bf{B}}\rb={\bf{0}}
\ee

\noindent or

\be\tag{28}
\frac{\partial A_i}{\partial t}+\frac{c}{n_e e}B_j\lp\partial_j B_i\rp=\frac{c}{n_e e}B_j\lp \partial_i B_j\rp.
\ee

In 2D\footnote{See Appendix for the 3D formulation in the massless-electron limit.}, where ${\bf A}=A\hat{\bf i}_z$, equation (28) becomes (upon normalizing $t$ appropriately),

\be\tag{29}
\frac{\partial A}{\partial t}=\lb B_z, A\rb.
\ee

 Equation (29) implies that the out-of-plane magnetic field $B_z$ plays the role of the stream function for the in-plane electron flow, as to be expected. \\
\indent Equation (29) has the Hamiltonian formulation,

\be\tag{30}
\mathscr H=\underset{\mathscr A}\int AB_z ~d\mathscr A.
\ee

The Hamilton's equation, on taking $A$ to be the canonical variable, is given by

\be\tag{31}
\frac{\partial A}{\partial t}~=\mathscr{J} ~\frac{\delta \mathscr H}{\delta A}
\ee

\noindent where $\mathscr J$ is an $A$-dependent differential operator which produces a skew-symmetric transformation of scalar functions vanishing on $\partial\mathscr A$ and satisfies a closure condition on an associated symplectic two-form,

\be\tag{32}
\mathscr J = \lb \cdot, A\rb.
\ee

\noindent Equation (31) is then just equation (29). \\
\indent The operator $\mathscr{J}$ induces a Poisson bracket as follows,

\be\tag{33}
\begin{aligned}
\lb F, G \rb& =\lp\frac{\delta F}{\delta A},~\mathscr{J}\frac{\delta G}{\delta A}\rp\\
\\
&= \underset{\mathscr A}\int \lp \frac{\delta F}{\delta A}\cdot \lb \frac{\delta G}{\delta A}, A \rb\rp d \mathscr A\\
\\
&= \underset{\mathscr A}\int\lp A\lb\frac{\delta F}{\delta A}, \frac{\delta G}{\delta A} \rb \rp d \mathscr A
\end{aligned}
\ee

\noindent which is a bilinear function defined on admissible functionals $F[A]$ and $G[A]$, satisfying the conditions, 

\be\tag{34}
\nabla \cdot
\begin{pmatrix}\delta F/\delta A\\\delta G/\delta A\end{pmatrix}
=0~\text{in}~ A ~\text{and}~|\frac{\delta F}{\delta A}|, |\frac{\delta G}{\delta A}| =0 ~\text{on}~\partial \mathscr A.
\ee

The Casimir invariants are then solutions of the equation,

\be\tag{35}
\mathscr J \frac{\delta\mathscr C}{\delta A}=0
\ee

\noindent which leads to

\be\tag{36}
\mathscr C=\int f\lp A\rp d \mathscr A
\ee

\noindent $f(A)$ being an arbitrary function.\\

Minimization of $\mathscr H$, keeping $\mathscr C$ fixed, leads to the Beltrami state,

\be\tag{37}
\frac{\delta \mathscr H}{\delta A}=\lambda \frac{\delta\mathscr C}{\delta A}
\ee

\noindent or 

\be\tag{38}
B_z=\lambda f^\prime\lp A\rp
\ee

\noindent which leads to,

\be\tag{39}
{\bf\mathscr J}=c\lambda f^{\prime\prime}\lp A\rp {\bf B}\equiv a{\bf B}
\ee

\noindent describing a force-free state in the massless-electron limit (contrary to the general EMHD case). So, EMHD, in the massless-electron limit, mimics usual MHD (this result holds in 3D as well - see Appendix). 
\vspace{.3in}

\noindent\Large\textbf{6. Discussion}\\

\large The intrinsic tendency of real plasmas (or fluids) toward Beltramization is a long standing issue awaiting full clarification. Work to date has indicated that the Beltramization process provides the means via which plasma systems can accomplish,
\begin{itemize}
  \item[*] an effective depletion of non-linearities,
  \item[*] ergodicity of the streamlines of the plasma flow in question (Moffatt \cite{Mof}),
  \item[*] selective dissipation of total energy (Woltjer \cite{Wol}).
\end{itemize}

In this paper, the Hamiltonian formulations along with the Poisson brackets for 2D EMHD flows are developed. These formulations are used to deduce the Beltrami states for 2D EMHD flows. Though there is no force-free Beltrami state in 2D EMHD flows, in the massless-electron limit Beltrami states become force-free. So, EMHD, in the massless-electron limit, mimics usual MHD (this result holds in 3D as well - see Appendix).

\vspace{.3in}

\noindent\Large\textbf{Acknowledgments}\\

\large Part of this work was carried out during the course of my participation in the Turbulence Workshop at the Kavli Institute of Theoretical Physics, Santa Barbara. I am thankful to Dr. Maxim Lyutikov for a helpful correspondence. This research was supported in part by NSF grant No. PHY05-51164. \\

\vspace{.3in}

\noindent\Large\textbf{Appendix}\\

\large In the massless electron limit, upon appropriately renormalizing $t$, equation (27) can be rewritten as

\be\tag{A$\cdot$1}
\frac{\partial {\bf B}}{\partial t}+\nabla\times\{\lp\nabla\times{\bf B}\rp\times{\bf B}\}={\bf 0}
\ee

Equation (A$\cdot$1) has the Hamiltonian characterization,

\be\tag{A$\cdot$2}
\mathscr H =\frac{1}{2}\underset{\mathscr V}\int{\bf B}^2 d\mathscr V
\ee

\noindent where $\mathscr{V}$ is the volume occupied by the plasma. Hamilton's equation, on taking $\textbf{B}$ to be the canonical variable, is given by

\be\tag{A$\cdot$3}
\frac{\partial {\bf B}}{\partial t}=\mathscr J \frac{\delta \mathscr H}{\delta {\bf B}}
\ee

\noindent where $\mathscr{J}$ is a $\textbf{B}$-dependent differential operator, which produces a skew-symmetric transformation of vector functions vanishing on $\partial \mathscr V$ and satisfies a closure condition on an associated symplectic two-form\footnote{(A$\cdot$4) is similar to the one given by Olver \cite{Olv} for incompressible hydrodynamics (with the flow vorticity $\mathbf\omega$ taking the place of ${\bf B}).$},

\be\tag{A$\cdot$4}
\mathscr J =\nabla\times\{{\bf B}\times\lp\nabla\times\lp\cdot\rp\rp\}.
\ee

\noindent Equation (A$\cdot$3) is then just equation (A$\cdot$1). \\

The operator $\mathscr{J}$ may be seen to induce the Poisson bracket (similar to the one given by Kuznetsov and Mikhailov \cite{KMik}),

\be\tag{A$\cdot$5}
\lb F, G \rb = \lp \frac{\delta F}{\delta {\bf B}},\mathscr J \frac{\delta G}{\delta {\bf B}} \rp
= \underset{\mathscr V}\int{\bf B}\cdot\lb\lp\nabla\times\frac{\delta G}{\delta {\bf B}}\rp\times\lp \nabla\times\frac{\delta F}{\delta {\bf B}} \rp\rb d \mathscr{V}
\ee

\noindent which is a bilinear operation defined on the admissible functionals satisfying the conditions, 

\be\notag
\nabla \cdot
\begin{pmatrix}\delta F/\delta {\bf B}\\\delta G/\delta {\bf B}\end{pmatrix}
={\bf 0}~\text{in}~ \mathscr V
\ee

\noindent and 

\be\tag{A$\cdot$6}
|\frac{\delta F}{\delta {\bf B}}|, |\frac{\delta G}{\delta {\bf B}}| =0 ~\text{on}~\partial \mathscr V.
\ee

The Casimir invariant is then given by the solution of the equation,

\be\tag{A$\cdot$7}
\mathscr J\frac{\delta\mathscr C}{\delta {\bf B}}={\bf 0}
\ee

\noindent which leads to the magnetic helicity,

\be\tag{A$\cdot$8}
\mathscr C = \underset{\mathscr V}\int \textbf{A} \cdot \textbf{B}~d\mathscr V .
\ee

Minimization of $\mathscr H$, keeping $\mathscr C$ fixed, leads to the Beltrami state,

\be\tag{A$\cdot$9}
\frac{\delta \mathscr H}{\delta \textbf{B}} = \lambda \frac{\delta \mathscr C}{\delta \textbf{B}}
\ee

\noindent or

\be\tag{A$\cdot$10}
\textbf{B} = \lambda \textbf{A}
\ee

\noindent which leads to

\be\tag{A$\cdot$11}
\textbf{J} = c \lambda \textbf{B}
\ee

\noindent describing again a force-free state in the massless-electron limit. So, EMHD, in the massless-electron limit, mimics usual MHD.

\end{document}